\documentclass{nature}

\usepackage{graphicx}
\usepackage{amsmath}
\usepackage{amssymb}
\usepackage{color}
\usepackage{dcolumn}
\usepackage{epsfig}
\usepackage{bm}

\title{Crossover from Metal to Insulator in Dense Lithium-Rich Compound $\textrm{CLi}_{4}$}

\author{Xilian Jin$^1$, Xiao-Jia Chen$^2$, Huadi Zhang$^1$, Quan Zhuang$^1$, Kuo Bao$^1$, Dawei Zhou$^3$, Qiang Zhou$^1$, Zhi He$^1$, Bingbing Liu$^1$, \& Tian Cui$^{1*}$}

\begin{document}

\maketitle

\begin{affiliations}
 \item State Key Laboratory of Superhard Materials, College of Physics, Jilin University, Changchun 130012, China
 \item Center for High Pressure Science and Technology Advanced Research, Shanghai 201203, China
 \item College of Physics and Electronic Engineering, Nanyang Normal University, NanYang 473061, China
\end{affiliations}

\begin{abstract}
Crystal structures of $\textrm{CLi}_4$ compounds are explored through \emph{ab} \emph{initio} evolutionary methodology. Phase transition from metal to semimetal and semiconductor, and eventually to insulator with increasing pressure are revealed under pressure. Pressure-induced evolution of anti-metallization has been described quantitatively by Fermi Surface Filling Parameter and electron energy band gap using \emph{ab} \emph{initio} \emph{GW} calculations. Anti-metallization is attributed to the hybrid valence electrons and their repulsion by core electrons into the lattice interstices. Very weak electron-phonon coupling interactions are found in the metallic phases, resulting in very low superconducting temperature.
\end{abstract}

Hydrogen is the most abundant element in the universe. Though, at the top of the alkali metal column in the periodic table, hydrogen is not an alkali metal and is an insulator in ordinary conditions. The pressure-induced transformation is expected to be accompanied by a decrease in the band gap and eventual formation of metallic hydrogen \cite{1-Wigner1935}, attracting a lot of interest. Lithium has a single valence electron similar to hydrogen, but lithium is a good conductor of heat and metal under ambient conditions. Unlike the metallization of hydrogen under high pressure, lithium can transform from a nearly free-electron metallic solid to an insulating one, i.e. pressure-induced anti-metallization \cite{2-Neaton1999}. As a contrasting counterpart, lithium can provide meaningful referential aspect for understanding of hydrogen at pressures.

Recently hydrogen-rich materials have been investigated extensively and profoundly. Alkali metals, e.g. lithium and potassium, together with hydrogen can be compressed into alkali high-hydride alloys \cite{3-Zurek2009, 4-Zhou2012}, and metallize at pressures much lower than that required for pure hydrogen due to the "chemical precompression". A special mention has been made on the group-IV hydrides, calculations and experiments on compressed silane \cite{5-Feng2006, 6-Kim, 7-Eremets2008}, disilane \cite{8-Jin2010, 9-Folres2012}, germane \cite{10-Chen2009}, etc. show the possibility of metallization at moderate pressures. These studies are believed to be useful to realize metallic hydrogen, which has been described as "the holy grail of high-pressure physics" \cite{11-Cornell1998}. Accordingly, the studies of group-IV lithium compounds should benefit to the comprehension of the metallization of group-IV hydrides which are very important for understanding and researching metallic hydrogen.

Lithium is well-known for its ability to form heterogeneous clusters with various elements. Since the experimental discovery and theoretical verification of "hyperlithiated" bonding, which involves formal violations of the octet rule in doped Li clusters \cite{12-Schleyer1982}, there have been lots of theoretical and experimental investigations with the purpose of insight into the specific structural and electronic properties of these systems. Such systems include $\textrm{CLi}_n$ \cite{13-Schleyer1983, 14-Kudo1992}, etc. CLi$_{6}$ was first predicted \cite{13-Schleyer1983} and later verified experimentally \cite{14-Kudo1992} to be stable with respect to dissociation into the stoichiometric CLi$_{4}$ and Li$_{2}$ molecules. To enhance the H$_{2}$ binding and attain high storage capacity, some nanostructures have been functionalized with polylithiated molecules, e.g. CLi$_{4}$ for hydrogen storage. The stabilities of the designed functional materials for H$_{2}$ storage have been verified by means of molecular dynamics simulations \cite{15-Er2009}.

In view of interests mentioned above, we have explored the crystal structures of lithium-rich compound CLi$_{4}$ in a wide pressure range from ambient pressure to 300 GPa. Contrary to the group-IV hydrides, a reverse pressure-induced metallization is found in the CLi$_{4}$ system under high pressure. We have found that the phase transitions from metallic \textit{P}1, \textit{P}-1, \textit{P}2$_{1}$, \textit{C}2/\textit{m}, to semimetal and semiconductor of \textit{R}-3\textit{m}, and eventually to insulated \textit{Fddd} with increasing pressure quantitatively by Fermi Surface Filling Parameter and electron energy band gaps using \emph{ab} \emph{initio} \emph{GW} calculations. Pressure-induced anti-metallization is found not only in the CLi$_{4}$ system, but also in the phase of \textit{R}-3\textit{m}. Unlike group-IV hydrides, the electron-phonon coupling interactions are fairly weak and the superconducting temperatures are very low under high pressure. Finally, we attribute anti-metallization to the hybrid valence electrons and their repulsion by core electrons into the lattice interstices.

\begin{figure}
\caption{\textbf{(Color online) Calculated enthalpy difference versus pressure for various structures.}  Enthalpies per CLi$_{4}$ unit of structures as functions of pressure with respect to the decomposition (C+4Li and (C$_{2}$Li$_{2}$)/2+3Li) enthalpies (black and red lines, respectively), which are calculated by adopting the C, Li and C$_{2}$Li$_{2}$ structures from references \cite{23-Katzke2006,24-Hanfland2000,25-Yao2009} and \cite{26-Ruschewitz1999,27-Chen2010,28-Nylen2012} at corresponding pressures.}
\end{figure}

The enthalpy-pressure curves of CLi$_{4}$ show the possible decomposition to the elements and reported compounds, see Figure 1, and the other possible lithium-rich carbides CLi$_{n}$ (n=2$\sim$6) and their stability can be found in supplementary material (see Extended Data Figure 1). Obviously, the compound of CLi$_{4}$ can be stable at least above 1 bar and below 300 GPa. We carry out unconstrained searches at fixed pressures of 1bar, and 5, 10, 50, 100, 200, 300 GPa using one, two, three, four, five, and six CLi$_{4}$ formula units (\emph{f.u.}) per cell, and have found many different relaxed structures. A special one with twelve units, i.e. sixty atoms per cell, is also performed at the low pressure of 1bar and 5 GPa, respectively. Figure 1 shows our calculated enthalpies of the candidate structures including USPEX and XY$_{4}$-type ones, together with the decomposition (C+4Li and (C$_{2}$Li$_{2} $)/2+3Li) enthalpies (black and red lines, respectively). Before about 3 GPa, CLi$_{4}$ is unstable and decomposes into a C$_{2}$Li$_{2}$ and Li mixture. After about 3 GPa, CLi$_{4}$ compounds become stable energetically over the decomposition enthalpies lines, as shown in Figure 1. Several different structures with space group of \textit{P}1 are found at about 5 GPa with very close enthalpies. So, CLi$_{4}$ compounds can be considered as amorphous at this low pressure range. To simplify the structure of amorphous phase, a small  cell with thirty atoms is chosen as an example for the \textit{P}1 structures shown in supplementary material (see Extended Data Figure 3. \textbf{a}), and will be discussed below. Details of two \textit{P}1 structures as shown in Extended Data Figure 3 are provided in the supplementary material (see Extended Data Table 3 and 4). The \textit{P}-1 phase becomes stable above 7 GPa, which is obviously energetically favored over the others until the pressure close to 18 GPa. At 18 GPa, \textit{P}-1 phase starts to transform into \textit{P}2$_{1}$ CLi$_{4}$. Then \textit{P}2$_{1}$ phase has the obviously lowest enthalpy among the structures until 40 GPa. After 40 GPa, \textit{P}2$_{1}$ phase transforms into \textit{C}2/\textit{m} phase. At 60 GPa, a high symmetry phase of \textit{R}-3\textit{m} appears and is stable until 128 GPa. The magnified images of enthalpy curves in this pressure range can be found in Extended Data Figure 2. A pressure-induced symmetrization processes from low symmetry phases to the high symmetry one with increasing pressure until 128 GPa is demonstrated. After 128 GPa, another high symmetry phase of \textit{Fddd} appears with 32 symmetry operators, which is slightly less than the 36 operators of \textit{R}-3\textit{m}. Obviously, the five phases, i.e., \textit{P}-1, \textit{P}2$_{1}$, \textit{C}2/\textit{m}, \textit{R}-3\textit{m} and \textit{Fddd}, are the most favored structures of CLi$_{4}$ compounds under high pressures.

The favored structures of CLi$_{4}$ obtained in this work are shown in Figure 2, and the corresponding parameters at respective pressures are listed in Extended Data Table 1. At low pressure range, the triclinic phase of \textit{P}-1 consists of two \emph{f.u.} of CLi$_{4}$ (i.e., C$_{2}$Li$_{8}$) with ten unequivalent atoms in the primitive cell, as shown in Figure 2 \textbf{a}. Each unequivalent atom occupies the crystallographic 2\textit{i} position and the site symmetry is \textbf{1}. The \textit{P}2$_{1}$ phase consists of two \emph{f.u.} of CLi$_{4}$ (i.e., C$_{2}$Li$_{8}$) with ten unequivalent atoms in a monoclinic crystal lattice, as shown in Figure 2 \textbf{b}. The eight Li atoms and two C atoms occupy the crystallographic 2\textit{a} position with the site symmetry of \textbf{1}. From the Figure 2 \textbf{c}, we can see that
\textit{C}2/\textit{m} CLi$_{4}$ is another monoclinic phase. Consisting of two \emph{f.u.} of CLi$_{4}$ (i.e., C$_{2}$Li$_{8}$) in the crystal lattice, there are three unequivalent atoms in the conventional cell. Two Li atoms occupy the crystallographic 4\textit{i} position with \textbf{\emph{m}} site symmetry, and one C atom locates at 2\textit{c} position with \textbf{2}/\textbf{\emph{m}} site symmetry. Lattice parameters of this monoclinic phase at 40 GPa are shown in Extended Data Table 1. The \textit{R}-3\textit{m} phase contains three \emph{f.u.} of CLi$_{4}$ (i.e., C$_{3}$Li$_{12}$) in a hexagonal crystal lattice, as shown in Figure 2 \textbf{d}. The corresponding lattice parameters of this hexagonal phase at 60 GPa are listed in Extended Data Table 1. There are three unequivalent atoms, including two Li atoms and one C atom occupying the crystallographic 6\textit{c} and 3\textit{b} positions with \textbf{3}\textbf{\emph{m}} and \textbf{-3}\textbf{\emph{m}} site symmetry, respectively. The \textit{Fddd} phase consists of eight \emph{f.u.} of CLi$_{4}$ (i.e., C$_{8}$Li$_{32}$) in an orthorhombic crystal lattice, as shown in Figure 2 \textbf{e}. The corresponding lattice parameters of this phase at 150 GPa are listed in Extended Data Table 1. There are two unequivalent atoms in the conventional cell. The Li atom occupies the crystallographic 32\textit{h} position with the site symmetry of \textbf{1}, and C atom locates at the 8\textit{a} site with the site symmetry of \textbf{222}, respectively.

\begin{figure}
\caption{\textbf{The structures of CLi$_4$ under high pressures.} \textbf{a}, The \textit{P}-1 phase at 15 GPa; \textbf{b}, \textit{P}2$_1$ phase at 30 GPa; \textbf{c}, \textit{C}2/\textit{m} phase at 40 GPa; \textbf{d}, \textit{R}-3\textit{m} phase at 60 GPa; \textbf{e}, \textit{Fddd} phase at 150 GPa. Black and blue atoms are C and Li, respectively.}
\end{figure}

After the discussion on the viewpoint of thermodynamics above, the studies focus on mechanics and the lattice dynamics to test the reasonableness of propositional structures. The mechanical stability provides a useful understanding for the structure of crystals. The strain energy of a crystal must be positive against any homogeneous elastic deformations, i.e., the matrix of elastic constants \textit{C}$_{ij}$ must be positive definite \cite{29-JF1985}. The elastic constants have been calculated and listed in Extended Data Table 5. Obviously, the elastic constants of the structures satisfy the mechanical stability criteria \cite{29-JF1985,30-Wu2007}, indicating that the five structures are mechanically stable. The charts of phonon band structure and the projected DOS of \textit{P}-1, \textit{P}2$_{1}$, \textit{C}2/\textit{m}, \textit{R}-3\textit{m}, and \textit{Fddd} phases calculated at selected pressures are shown in Extended Data Figure 4. The absence of imaginary frequency modes indicates that these structures are stable dynamically. A direct result from PVDOS of these structures shows that the lighter Li atoms contribute more in frequency vibrational modes than the heavier C atoms.

\begin{figure}
\caption{\textbf{(Color online) Pressure-induced evolutions of anti-metallization and superconductivity.} The evolution from metal to insulator described quantitatively with Fermi Surface Filling Parameter (before 70 GPa, pink dash line and olive solid line in left part) and Electron Energy Band Gap (after 70 GPa, olive and purple solid lines in right part)using \emph{ab initio GW} calculations. \textbf{a-e}, The 3D FS are shown as upside calculated by \emph{GW} method for the corresponding phases at selected pressures. \textbf{f-g}, 3D FS of $R$-3$m$ CLi$_4$ using \emph{ab initio} DFT calculation, and its projected FS nesting function $\xi(Q)$ along high-symmetry directions at 60 GPa. The superconducting $T$$_c$ and electron-phonon coupling parameter $\lambda$ magnified by ten times as functions of pressure are depicted with red and navy solid line in left party, respectively.}
\end{figure}

It is known that \emph{ab} \emph{initio} DFT calculation usually gives incorrect estimate of fundamental energy band gap of materials \cite{6-Kim,31-G2002}. Recently, \emph{ab} \emph{initio} \emph{GW} calculations show good improvement in predicting energy band gap. The \textit{GW} correction can yield a reliable band structure of candidates to ensure that they are indeed metallic under pressure. In view of this, it is worth to perform electronic band structure calculations at the level of \emph{ab} \emph{initio} \emph{GW} method. From the insets \textbf{e} and \textbf{f} of 3DFS of \textit{R}-3\textit{m} at 60 GPa in Figure 3, we can clearly find the difference between \emph{ab} \emph{initio} DFT calculation and \emph{ab} \emph{initio} \emph{GW} method.

To investigate the degree of metallization quantitatively in metal CLi$_{4}$ under pressure, Fermi Surface (FS) Filling Parameter $\zeta $ is used, where $\zeta $ is described as $\zeta=\frac{1}{N}\sum_{kn}\delta(\varepsilon_{kn}-\varepsilon_{F})\propto \frac{N(0)}{C\sqrt{\varepsilon_F}}$. Obviously, this parameter can be deduced by the nesting function $\xi \left( Q\right)$ \cite{32-Kasinathan2006}. Here \textit{N}$\left(0\right) $ is the FS density of states, and \textit{C}$\sqrt{\varepsilon_{_{F}}}$ accounts for the intersection point between envelope and Fermi line in the chart of electron density of state of metal. This parameter can be used to describe free electrons ratio in metal. Obviously, $\zeta $ is the maximum value of $\xi \left( Q\right) $ at Gamma point and means the degree of metallization, as shown in inset \textbf{g} of Figure 3.

A remarkable property of CLi$_{4}$ is the existence of phase transitions from metal to insulator with increasing pressures. At the lower pressure range below 70 GP, \textit{P}1, \textit{P}-1, \textit{P}2$_{1}$, \textit{C}2\textit{m}, and \textit{R}-3\textit{m} phases are all metallic, and the degree of metallization declines gradually with increasing pressure except a slight pressure-induced metallization from phase \textit{P}-1 to \textit{P}2$_{1}$, as described in Figure 3 (left part). At about 70 GPa, the metallic character of \textit{R}-3\textit{m} CLi$_{4}$ disappears, and the non-metallic character of energy band gap begins to appear. Above 70 GPa, the energy band gap of \textit{R}-3\textit{m} phase grows bigger and bigger until reaching a value of 2.43 eV at 128 GPa with \emph{GW} corrections, as shown in Figure 3 (right part). So, the \textit{R}-3\textit{m} CLi$_{4}$ is a transitional phase from metallic property to non-metallic one. After 128 GPa, insulated \textit{Fddd} phase appears with a band gap value of 3.26 eV at 128 GPa. With further increase of pressure, the band gap has a slight elevation until 300 GPa. So, the decrease of metallization and increase of non-metallic character are revealed in compressed lithium-rich compound CLi$_{4}$ system, as described quantitatively by the FS Filling Parameter $\zeta $ and electron energy band gaps.

\begin{figure}
\caption{\textbf{(Color online) Electronic structures of the CLi$_4$ phases at selected pressures.} Blue and black atoms represent Li and C, respectively. \textbf{a}, $P$2$_1$ phase at 30G. Left up and down are the PDOS (partial DOS)  and integrated PDOS of element Li and C. Meddle is 3D-ELF (3 dimensional electron localization function) map with the elf value of 0.5, and right is the corresponding 2D-ELF (2 dimensional electron localization function) slice along the (001) plane. \textbf{b}, $Fddd$ phase at 150G. Left are PDOS and integrated PDOS of element Li and C. Meddle represents 3D-ELF with the elf value of 0.5, and right depicts corresponding 2D-ELF for the (010) plane.}
\end{figure}

Subsequently, calculations about electronic structures have been performed to reveal the emergence of anti-metallization in the dense lithium-rich compound CLi$_{4}$. Hybridizations of valence electrons (Li-s, Li-p, and C-2p) have been discovered under pressure, see PDOS in Figure 4 (left part). At lower pressure range, the interaction between valence electrons and ionic cores is weak, and the nonlocal electrons near Fermi energy account for the metallic property, as displayed in Figure 4 \textbf{a} (middle and right). From 3D-ELF and 2D-ELF in Figure 4 \textbf{a}, we can see that the conductivity comes from connected regions where the value of ELF is about 0.5. But under higher pressure, volume contraction in crystal lattice increasingly shortens interatomic distances, and hybrid electrons are repulsed by the core electrons into the lattice interstices, see Figure 4 \textbf{b} (middle and right). From 3D-ELF and 2D-ELF in Figure 4 \textbf{b}, we can see that nearly free electrons with the ELF value of 0.5 become totally localized and are confined to disconnected areas in interstitial positions of atoms. As a direct result of charge distributions under pressure, charge transfer has been observed, see Extended Data Table 6. With the increased overlap of the valence wavefunctions with pressure, the s-state electrons transfer to the p-state electrons in element Li by compression. Similar charge transfers from s-state to p-state have been reported in pure lithium under high pressure \cite{33-Boettger1989}. Meanwhile, charge transfers from C-2s to C-2p, and partially to Li-2p have been observed by the integration of PDOS. So, high pressure induces the decreased electronegativity of element C and increased the one of element Li.

Finally, superconductivities are investigated under the framework of BCS theory, as show in Figure 3 (left part). The electron-phonon coupling parameter (EPC $\lambda $), the logarithmic average phonon frequency $(\omega_{\log})$, and the Eliashberg phonon spectral function $(\alpha^2F(\omega))$ are calculated at high pressures. The resulting EPC $\lambda$ for \textit{P}1, \textit{P}-1, \textit{P}2$_{1}$, \textit{C}2/\textit{m}, and \textit{R}-3\textit{m} phases at 5, 15, 30, 40, 50, 60 GPa are 0.533, 0.398, 0.425, 0.364, 0.276, and 0.212, respectively, indicating that the EPC is very weak. The superconducting critical temperature can be estimated from the Allen--Dynes modified McMillan equation \textit{T}$_{c}=\frac{\omega _{\log }}{1.2}\exp \left[ \frac{1.04\left( 1+\lambda \right) }{\lambda -\mu ^{\ast }\left( 1-0.62\lambda \right) }\right]$ \cite{34-Allen1975}, which has been found to be highly accurate for many materials with $\lambda <1.5$. The $\omega _{\log }$ can be calculated directly from the phonon spectrum, and the values for the metallic phases mentioned above, are 305.75, 416.16, 400.64, 405.86, 484.95, and 560.85 K at selected pressures, respectively. The Coulomb pseudopotential $\mu ^{\ast }$ is often taken as 0.1 for most metals. Using $\mu ^{\ast }$ of 0.1, the estimated \textit{T}$_{c}$ are 4.725, 1.691, 2.344, 0.949, 0.094, and 0.001 K, respectively, revealing a similar varied tendency with anti-metallization under high pressure.

The dens lithium-rich compound CLi$_{4}$ demonstrates an unexpected process of anti-metallization, which is completely contrary to their counterpart of group-IV hydrides. Pressure-induced anti-metallization such as reported here also is expected in pure elements and compounds, such as alkali metal \cite{2-Neaton1999, 35-Ma2009} and some alkali metal compounds, where the interaction between valence electrons and ionic cores is very strong. During metallizing or anti-metallizing in metallic states, Fermi Surface Filling Parameter (FSFP) will be a valuable parameter to quantify the evolution of the free electrons, as described here.

\subsection{METHODS SUMMARY}

\subsection{}
\hspace{-3mm}
\emph{Ab} \emph{initio} evolutionary algorithm designed to search for the structure with the lowest energy at a given pressure has been implemented in USPEX code \cite{16-Oganov2006}. The underlying \emph{ab} \emph{initio} structure relaxations have been performed using density functional theory within the Perdew-Burke-Ernzerhof (PBE) parameterization of the generalized gradient approximation (GGA) \cite{17-Perdew1996} as implemented in the Vienna \emph{ab} \emph{initio} simulation package VASP code \cite{18-Kresse1996}. The all-electron projector-augmented wave (PAW) method \cite{19-Blochl1994} is adopted with the PAW potentials taken from the VASP library, where 1s2s2p and 2s2p are treated as valence electron configurations for Li and C atoms, respectively. A plane-wave basis set with an energy cutoff of 1000 eV is used and gives well converged total energies.

The plane-wave pseudopotential method within the PBE-GGA, through the Quantum-ESPRESSO (QE) package \cite{20-Giannozzi2009} is employed to study the electronic properties, lattice dynamics, and electron-phonon coupling (EPC). In the QE code, the favored phases from PAW calculations are fully reoptimized within a force and energy convergence threshold of 10$^{-5}$ Ry/bohr and 10$^{-6}$ Ry to minimize the internal force and energy, respectively. The Troullier-Martins norm-conserving pseudopotentials for Li and C are generated using the FHI98PP code \cite{21-Fuchs1999} with 2s$^{1}$ and 2s$^{2}$2p$^{2}$ as valence electrons, respectively. The pseudopotentials are then carefully tested by comparing the calculated lattice parameter and electronic band structure with VASP codes. Convergence test gives the choice of kinetic energy cutoffs of 80 Ry, and the Monkhorst-Pack (MP) \cite{22-Monkhorst1976} grids of k-point sampling for each favored phases using a grid of spacing $2\pi \times 0.025 \mathrm{{\AA}}^{-1}$ in the Brillouin zone (BZ). A density of q mesh with $2\pi \times 0.04 \mathrm{{\AA}}^{-1}$ in the first BZ is used in the interpolation of the force constants for the phonon dispersion curve calculations, and a denser mesh of $2\pi \times 0.025 \mathrm{{\AA}}^{-1}$ is used for the phonon density of state (PDOS) curve calculations. Subsequently, EPC are calculated in the first BZ on the same MP q-point meshes using individual EPC matrices obtained with a grid of spacing $2\pi \times 0.025 \mathrm{{\AA}}^{-1}$. All the convergences of the plane-wave basis set and MP sampling are carefully examined by employing higher kinetic energy cutoffs and denser grids sets.

\vspace{5mm}
\subsection{Online Content}
Supplementary material is available in the online version of the paper.

\begin{addendum}

 \item
 This work was supported by the National Basic Research Program of China (No. 2011CB808200), Program for Changjiang Scholars and Innovative Research Team in University (No. IRT1132), the National Natural Science Foundation of China (Nos. 51032001, 11074090, 10979001, 51025206, 11174102), and National Found for Fostering Talents of basic Science (No. J1103202), the HeNan Joint Funds of the National Natural Science Foundation of China (No. U1304612 ), and the Special Funding of National Natural Science Foundation of China for Theoretical Physics ( No. 11247222 ). Parts of calculations are performed at the High Performance Computing Center (HPCC) of Jilin University.

 \item[Author Contributions]
 Xilian Jin, Xiao-Jia Chen, and Tian Cui contributed equally.

 \item[Author Information]
 Correspondence and requests for materials should be addressed to T. C. (cuitian@jlu.edu.cn).
\end{addendum}

\newpage
\vspace{8cm}
\begin{center}
\includegraphics[width=\columnwidth]{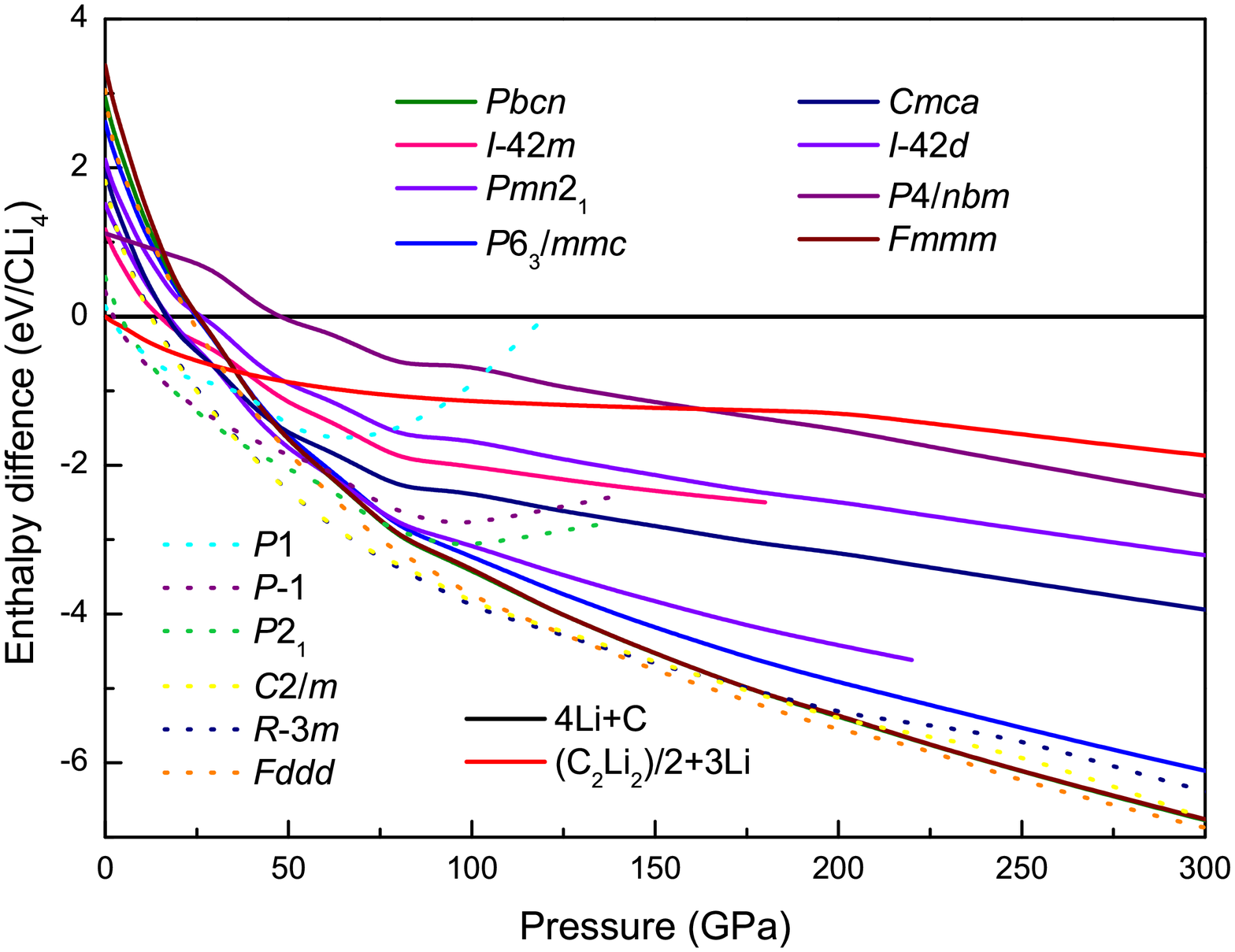}
{\item Figure 1.}
\end{center}

\newpage
\vspace{8cm}
\begin{center}
\includegraphics[width=\columnwidth]{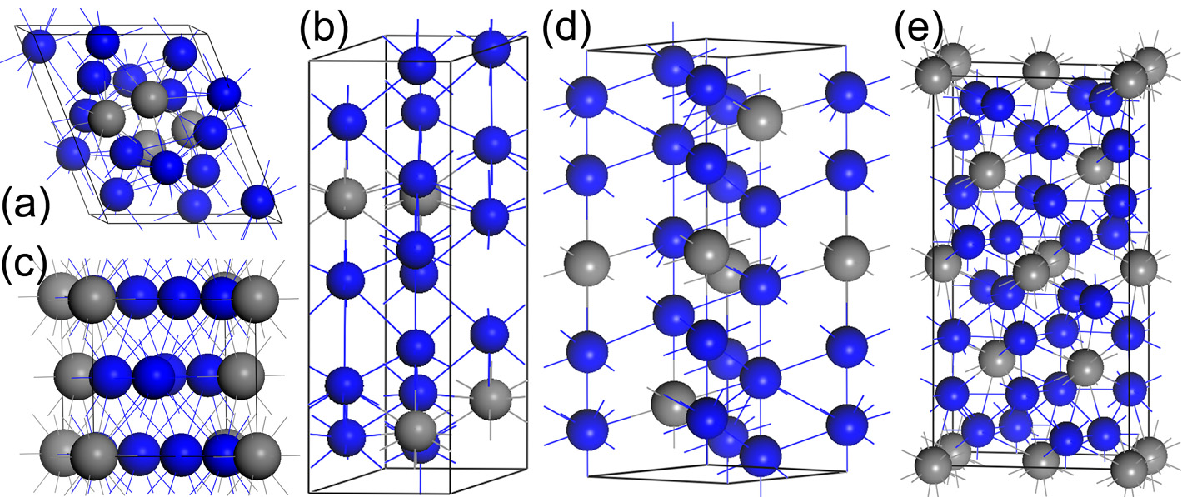}
{\item Figure 2.}
\end{center}

\newpage
\vspace{8cm}
\begin{center}
\includegraphics[width=\columnwidth]{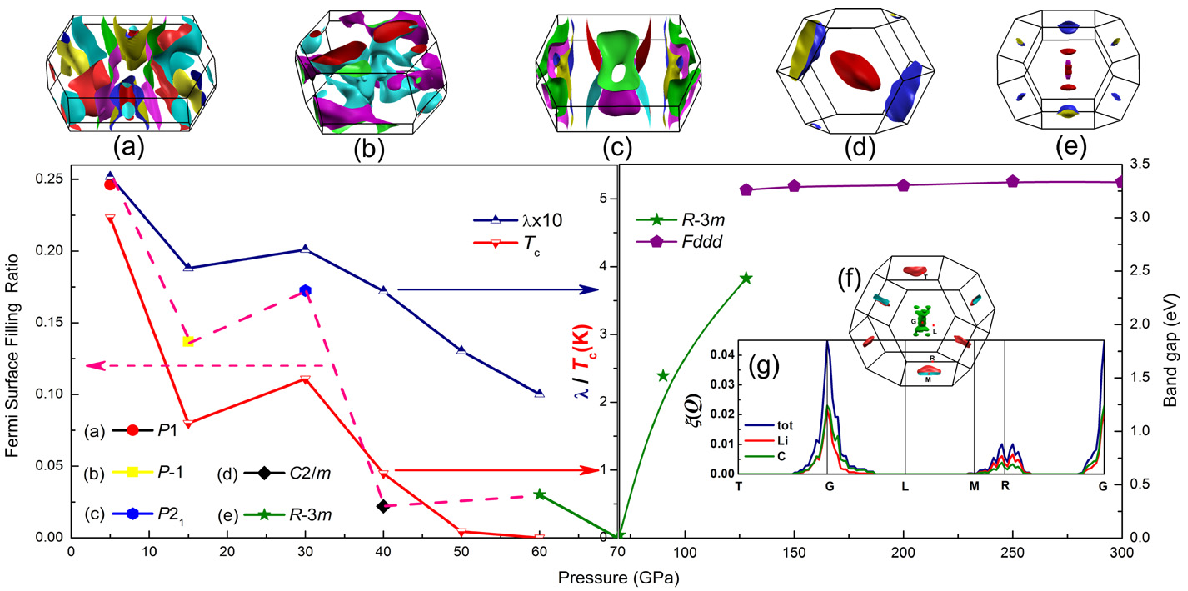}
{\item Figure 3.}
\end{center}

\newpage
\vspace{8cm}
\begin{center}
\includegraphics[width=\columnwidth]{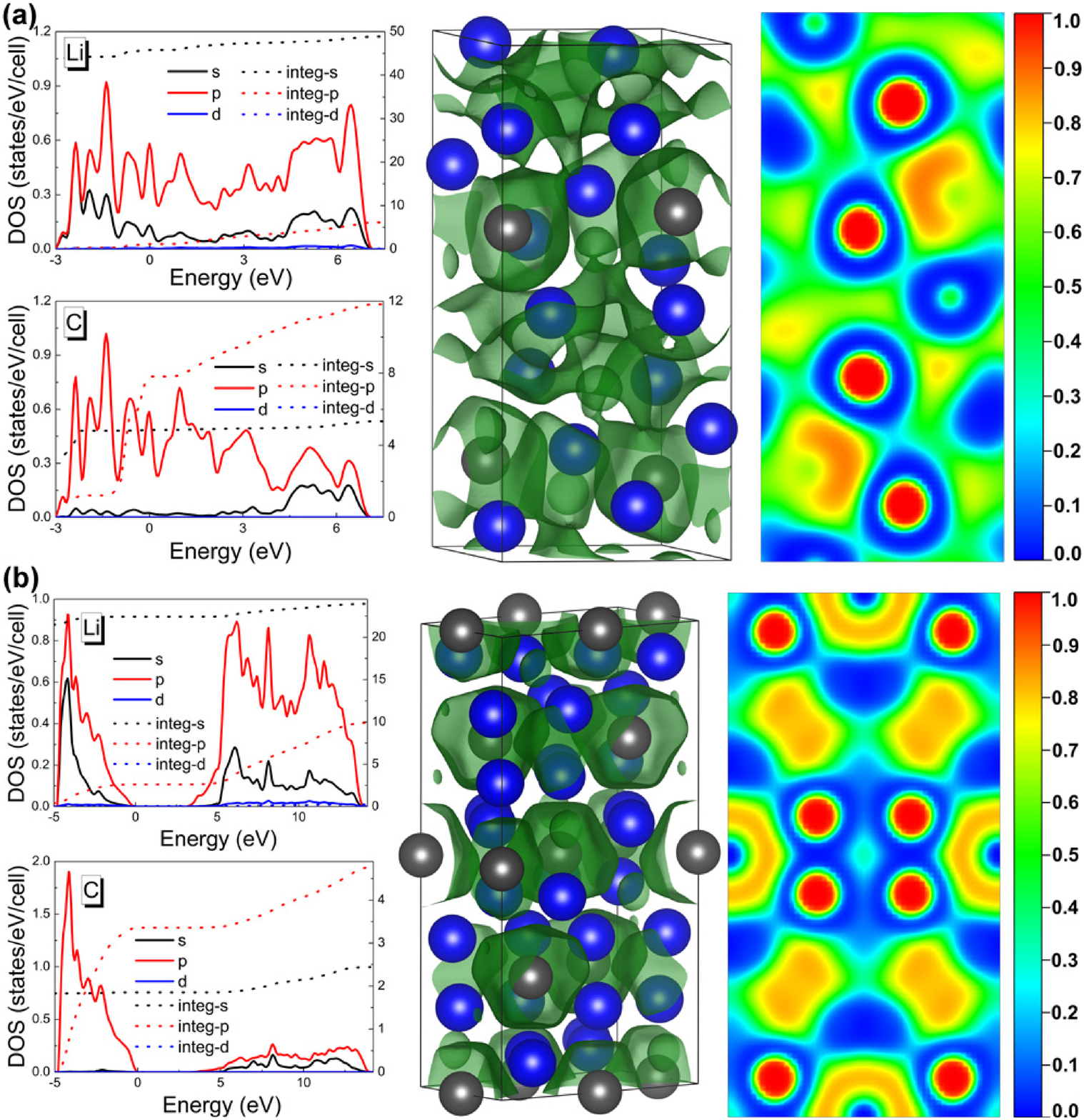}
{\item Figure 4.}
\end{center}

\end{document}